\documentstyle[12pt]{article}
\newlength{\extraspace}
\newlength{\extraspaces}
\newcommand{\be}{\begin{equation}
\addtolength{\abovedisplayskip}{\extraspaces}
\addtolength{\belowdisplayskip}{\extraspaces}
\addtolength{\abovedisplayshortskip}{\extraspace}
\addtolength{\belowdisplayshortskip}{\extraspace}}
\newcommand{\ee}{\end{equation}}
\newcommand{\ba}{\begin{eqnarray}
\addtolength{\abovedisplayskip}{\extraspaces}
\addtolength{\belowdisplayskip}{\extraspaces}
\addtolength{\abovedisplayshortskip}{\extraspace}
\addtolength{\belowdisplayshortskip}{\extraspace}}
\newcommand{\ea}{\end{eqnarray}}
\newcommand{\nonu}{\nonumber \\[.5mm]}
\newcommand{\A}{&\!\!\!}
\begin{document}
\thispagestyle{empty}
\begin{flushright}
SIT-LP-07/06 \\
{\tt hep-th/yymmnnn} \\
June, 2007
\end{flushright}
\vspace{7mm}
\begin{center}
{\large{\bf  $N = 2$ Supersymmetric QED equivalence \\[2mm]
of $N = 2$ Volkov-Akulov model}} \\[20mm]
{\sc Kazunari Shima}
\footnote{
\tt e-mail: shima@sit.ac.jp} \ 
and \ 
{\sc Motomu Tsuda}
\footnote{
\tt e-mail: tsuda@sit.ac.jp} 
\\[5mm]
{\it Laboratory of Physics, 
Saitama Institute of Technology \\
Fukaya, Saitama 369-0293, Japan} \\[20mm]
\begin{abstract}
We show explicitly in two dimensional spacetime ($d = 2$) 
that the $N = 2$ Volkov-Akulov model 
is equivalent to the spontaneously broken linear supersymmetry (LSUSY) 
interacting gauge theory for $N = 2$ vector and $N = 2$ scalar supermultiplets. 
The local gauge interaction of LSUSY is induced by the specific composite structure 
of the auxiliary fields and the consequent transformations.
%
\end{abstract}
\end{center}

\newpage

\noindent

Supersymmetry (SUSY) is a promising notion to search some unified theories 
beyond the Standard Model, 
which is realized as linear (L) representations or as nonlinear (NL) ones. 
The relation between the L and NL representations of SUSY 
is derived by linearizing of NLSUSY Volkov-Akulov (VA) model \cite{VA} 
which describes the dynamics of Nambu-Goldstone (NG) fermions \cite{SS}-\cite{O} 
indicating spontaneous supersymmetry breaking (SSB). 
Indeed, by means of the linearization of NLSUSY \cite{IK}-\cite{ST1} 
the VA model is equivalent to the various renormalizable L supermultiplet 
\cite{WZ1}-\cite{Fa} with the Fayet-Iliopoulos (FI) terms 
which lead to the SUSY breaking mass (gaps). 
The components of the equivalent LSUSY theories to the VA model 
are expanded as composites of the NG fermions according to SUSY invariant relations. 

Recently, we have shown \cite{ST2} the vanishing of the ordinary LSUSY 
Yukawa interaction terms (and mass terms) in NLSUSY representation 
by means of the SUSY invariant relations, 
namely, we have found that these terms can be added to a LSUSY free action 
without violating the equivalence between the VA model and the LSUSY theory. 
From the viewpoint of NLSUSY general relativity (NLSUSYGR) \cite{KS,ST3}, 
this result sheds light on the way to discuss the mediation of the mass generation 
in the SSB from a spacetime origin in NLSUSY 
to the (Yukawa-)Higgs origin in LSUSY \cite{ST4}. 
Because NLSUSYGR in asymptotic flat spacetime is essentially 
the VA model which is equivalent to the LSUSY theory with the mass gaps 
depending on the cosmological constant as well as the Newton gravitational constant. 
These arguments suggest that gauge interaction terms 
can be also discussed in the framework of NLSUSYGR through the linearization of NLSUSY. 

Therefore, we study in this paper the gauge interaction terms 
in the linearization of NLSUSY and show the SUSY QED equivalence of VA model. 
We consider the (realistic) $N = 2$ NLSUSY VA model 
and linearize it in $d = 2$ for simplicity of calculations 
by means of SUSY invariant relations 
to a (free) LSUSY theory which is described 
by minimally provided $SO(2)$ vector and scalar supermultiplets. 
The LSUSY and local $U(1)$ gauge invariant action is constructed in the LSUSY theory 
and the explicit form of the gauge interaction terms in NLSUSY representation 
is derived, which becomes nonvanishing four fermion self-interaction terms 
in contrast with the vanishing Yukawa interactions 
under the SUSY invariant relations. 
Furthermore we redefine the original SUSY invariant relations 
by adding terms depending on a gauge coupling constant 
such that the nontrivial four fermion self-interaction terms 
corresponding to the gauge interaction terms 
are cancelled out from the LSUSY and gauge invariant action in NLSUSY representation. 
According to the new definition of the SUSY invariant relations, 
we show explicitly that the $N = 2$ VA model is equivalent to the spontaneously 
broken LSUSY gauge theory containing the gauge interaction terms. 

Let us first introduce the $N = 2$ NLSUSY VA model \cite{VA} in terms of 
two (Majorana) NG fermions $\psi^i$ ($i = 1, 2$). 
\footnote{
Minkowski spacetime indices in $D = 2$ are denoted by $a, b, \cdots = 0, 1$ 
and $SO(2)$ internal indices are $i, j, \cdots = 1, 2$. 
The Minkowski spacetime metric is 
${1 \over 2}\{ \gamma^a, \gamma^b \} = \eta^{ab} = {\rm diag}(+, -)$ 
and $\sigma^{ab} = {i \over 2}[\gamma^a, \gamma^b] 
= i \epsilon^{ab} \gamma_5$ $(\epsilon^{01} = 1 = - \epsilon_{01})$, 
where we use the $\gamma$ matrices defined as $\gamma^0 = \sigma^2$, 
$\gamma^1 = i \sigma^1$, $\gamma_5 = \gamma^0 \gamma^1 = \sigma^3$ 
with $\sigma^I (I = 1, 2, 3)$ being Pauli matrices. 
}
The VA action for $\psi^i$ in $d = 2$ is written as the following spacetime volume form, 
\ba
S_{\rm VA} = - {1 \over {2 \kappa^2}} \int d^2 x \ \vert w \vert, 
\label{VAaction}
\ea
where $\kappa$ is a constant whose dimension is $({\rm mass})^{-1}$ and 
\be
\vert w \vert = \det(w^a{}_b) = \det(\delta^a_b + t^a{}_b), 
\ \ \ t^a{}_b = - i \kappa^2 \bar\psi^i \gamma^a \partial_b \psi^i. 
\ee
Eq.(\ref{VAaction}) is expanded in terms of $t^a{}_b$ or $\psi^i$ as 
\ba
S_{\rm VA} 
= \A \A - {1 \over {2 \kappa^2}} \int d^2 x 
\left\{ 1 + t^a{}_a + {1 \over 2!}(t^a{}_a t^b{}_b - t^a{}_b t^b{}_a) 
\right\} 
\nonu
= \A \A - {1 \over {2 \kappa^2}} \int d^2 x 
\left\{ 1 - i \kappa^2 \bar\psi^i \!\!\not\!\partial \psi^i 
- {1 \over 2} \kappa^4 
( \bar\psi^i \!\!\not\!\partial \psi^i \bar\psi^j \!\!\not\!\partial \psi^j 
- \bar\psi^i \gamma^a \partial_b \psi^i \bar\psi^j \gamma^b \partial_a \psi^j ) 
\right\} 
\nonu
= \A \A - {1 \over {2 \kappa^2}} \int d^2 x 
\left\{ 1 - i \kappa^2 \bar\psi^i \!\!\not\!\partial \psi^i 
\right. 
\nonu
\A \A 
\left. 
- {1 \over 2} \kappa^4 \epsilon^{ab} 
( \bar\psi^i \psi^j \partial_a \bar\psi^i \gamma_5 \partial_b \psi^j 
+ \bar\psi^i \gamma_5 \psi^j \partial_a \bar\psi^i \partial_b \psi^j ) 
\right\}, 
\label{VAaction2}
\ea
with the second term, 
$-{1 \over {2 \kappa^2}} t^a{}_a = {i \over 2} 
\bar\psi^i \!\!\not\! \partial \psi^i$, 
being the kinetic term for $\psi^i$. 
The VA action (\ref{VAaction}) is invariant (becomes a surface term) 
under NLSUSY transformations of $\psi^i$ 
parametrized by constant (Majorana) spinor parameters $\zeta^i$, 
\be
\delta_\zeta \psi^i = {1 \over \kappa} \zeta^i 
- i \kappa \bar\zeta^j \gamma^a \psi^j \partial_a \psi^i, 
\label{NLSUSY}
\ee
which correspond to the supertranslations of $\psi^i$ 
and Minkowski coordinates $x^a$. 
Eq.(\ref{NLSUSY}) satisfies a closed off-shell commutator algebra, 
\be
[ \delta_{Q1}, \delta_{Q2} ] = \delta_P(\Xi^a), 
\label{N2D2comm}
\ee
where $\delta_P(\Xi^a)$ means a translation with a generator 
$\Xi^a = 2 i \bar\zeta_1^i \gamma^a \zeta_2^i$. 

In order to construct a LSUSY action with gauge interaction terms in $N = 2$, $d = 2$, 
we next minimally provide $SO(2)$ vector and scalar supermultiplets, ${\bf V}$ and ${\bf \Phi}$. 
The $N = 2$ LSUSY free action for $N = 2$ off-shell vector supermultiplet ${\bf V}$ 
with a FI term is given by \cite{ST2} 
\be
S_{V0} = \int d^2 x \left\{ - {1 \over 4} (F_{ab})^2 
+ {i \over 2} \bar\lambda^i \!\!\not\!\partial \lambda^i 
+ {1 \over 2} (\partial_a A)^2 
+ {1 \over 2} (\partial_a \phi)^2 
+ {1 \over 2} D^2 
- {1 \over \kappa} \xi D \right\}, 
\label{Vaction}
\ee
where the component fields $(v^a, \lambda^i, A, \phi, D)$ 
($F_{ab} = \partial_a v_b - \partial_b v_a$) 
mean a $U(1)$ vector field, doublet (Majorana) fermions 
and a scalar field in addition to another scalar field 
and an auxiliary scalar field, respectively. 
Note that the off-shell fermionic and bosonic degrees of freedom 
of these component fields are balanced as 4 = 4. 
In Eq.(\ref{Vaction}) the last FI term with the real parameter $\xi$ 
gives the vacuum expectation value $<D> = {\xi \over \kappa}$ indicating SSB. 
The free action (\ref{Vaction}) is invariant under $N = 2$ LSUSY transformations 
parametrized by $\zeta^i$, 
\ba
\A \A 
\delta_\zeta v^a = - i \epsilon^{ij} \bar\zeta^i \gamma^a \lambda^j, 
\nonu
\A \A 
\delta_\zeta \lambda^i 
= (D - i \!\!\not\!\partial A) \zeta^i 
+ {1 \over 2} \epsilon^{ab} \epsilon^{ij} F_{ab} \gamma_5 \zeta^j 
- i \epsilon^{ij} \gamma_5 \!\!\not\!\partial \phi \zeta^j, 
\nonu
\A \A 
\delta_\zeta A = \bar\zeta^i \lambda^i, 
\nonu
\A \A 
\delta_\zeta \phi = - \epsilon^{ij} \bar\zeta^i \gamma_5 \lambda^j, 
\nonu
\A \A 
\delta_\zeta D = - i \bar\zeta^i \!\!\not\!\partial \lambda^i. 
\label{VLSUSY}
\ea
Eqs.(\ref{VLSUSY}) satify the following closed off-shell commutator algebra, 
\be
[ \delta_{Q1}, \delta_{Q2} ] = \delta_P(\Xi^a) + \delta_g(\theta), 
\label{N2D2commg}
\ee
where $\delta_g(\theta)$ is the $U(1)$ gauge transformation only for $v^a$ 
with a generator $\theta = - 2 (i \bar\zeta_1^i \gamma^a \zeta_2^i$ $v_a 
- \epsilon^{ij} \bar\zeta_1^i \zeta_2^j A 
- \bar\zeta_1^i \gamma_5 \zeta_2^i \phi)$. 

On the other hand, for $N = 2$ off-shell scalar supermultiplet ${\bf \Phi}$ 
we take ($\chi$, $B^i$, $\nu$, $F^i$), 
where $(\chi, \nu)$ for two (Majorana) fermions, 
$B^i$ for doublet scalar fields and $F^i$ for auxiliary scalar fields. 
$N = 2$ LSUSY free action for ${\bf \Phi}$ including FI terms 
with the real $SO(2)$ ($U(1)$) parameters $\xi^i$ is given by 
\be
S_{\Phi 0} = \int d^2 x \left\{ {i \over 2} \bar\chi \!\!\not\!\partial \chi 
+ {1 \over 2} (\partial_a B^i)^2 
+ {i \over 2} \bar\nu \!\!\not\!\partial \nu 
+ {1 \over 2} (F^i)^2 
- {1 \over \kappa} \xi^i F^i \right\}, 
\label{Saction}
\ee
which is invariant under the following $N = 2$ LSUSY transformations, 
\ba
\A \A 
\delta_\zeta \chi = ( F^i - i \!\!\not\!\partial B^i ) \zeta^i, 
\nonu
\A \A 
\delta_\zeta B^i = \bar\zeta^i \chi - \epsilon^{ij} \bar\zeta^j \nu, 
\nonu
\A \A 
\delta_\zeta \nu = \epsilon^{ij} ( F^i + i \!\!\not\!\partial B^i ) \zeta^j, 
\nonu
\A \A 
\delta_\zeta F^i = - i \bar\zeta^i \!\!\not\!\partial \chi 
- i \epsilon^{ij} \bar\zeta^j \!\!\not\!\partial \nu. 
\label{SLSUSY}
\ea
Eqs.(\ref{SLSUSY}) also satify the closed off-shell commutator algebra 
of Eq.(\ref{N2D2comm}). 

Here we discuss the relation between the NLSUSY VA model and the LSUSY theories 
described by the free actions (\ref{Vaction}) and (\ref{Saction}) 
for ${\bf V}$ and ${\bf \Phi}$. 
In the linearization of $N = 2$ NLSUSY, 
SUSY invariant relations between the NG fermions $\psi^i$ 
and the component fields in Eqs.(\ref{Vaction}) and (\ref{Saction}) 
are constructed such that the NLSUSY transformations (\ref{NLSUSY}) 
reproduce the LSUSY ones of Eqs.(\ref{VLSUSY}) and Eqs.(\ref{SLSUSY}). 
Indeed, for ${\bf V}$ the $SO(2)$ and $N = 2$ SUSY invariant relations 
of $(v^a, \lambda^i, A, \phi, D)$ as composites of $\psi^i$ 
are given in all orders as follows \cite{ST2}: 
\ba
\A \A 
v^a = - {i \over 2} \xi \kappa \epsilon^{ij} 
\bar\psi^i \gamma^a \psi^j \vert w \vert, 
\nonu
\A \A 
\lambda^i = \xi \left[ \psi^i \vert w \vert 
- {i \over 2} \kappa^2 \partial_a 
\{ \gamma^a \psi^i \bar\psi^j \psi^j 
(1 - i \kappa^2 \bar\psi^k \!\!\not\!\partial \psi^k) \} \right], 
\nonu
\A \A 
A = {1 \over 2} \xi \kappa \bar\psi^i \psi^i \vert w \vert, 
\nonu
\A \A 
\phi = - {1 \over 2} \xi \kappa \epsilon^{ij} \bar\psi^i \gamma_5 \psi^j 
\vert w \vert, 
\nonu
\A \A 
D = {\xi \over \kappa} \vert w \vert 
- {1 \over 8} \xi \kappa^3 
\Box ( \bar\psi^i \psi^i \bar\psi^j \psi^j ). 
\label{VSUSYinv}
\ea
In Eqs.(\ref{VSUSYinv}) the transformation of $v^a(\psi)$ 
under Eq.(\ref{NLSUSY}) gives the $U(1)$ gauge transformation 
besides the ordinary LSUSY one (\ref{VLSUSY}) as 
\be
\delta_\zeta v^a(\psi) 
= - i \epsilon^{ij} \bar\zeta^i \gamma^a \lambda^j(\psi) + \partial^a W(\zeta; \psi) 
\label{NLSUSYv}
\ee
with the $U(1)$ gauge transformation parameter, 
\be
W(\zeta; \psi) 
= \xi \kappa^2 \epsilon^{ij} \bar\zeta^i \psi^j \bar\psi^k \psi^k 
(1 - i \kappa^2 \bar\psi^l \!\!\not\!\partial \psi^l ). 
\label{W}
\ee
From Eq.(\ref{NLSUSYv}) the commutator algebra on $v^a(\psi)$ 
under Eq.(\ref{NLSUSY}) does not induce the $U(1)$ gauge transformation term 
$\delta_g(\theta)$ in Eq.(\ref{N2D2commg}) \cite{STT2,ST2}; 
namely, it is closed as Eq.(\ref{N2D2comm}). 
While, for ${\bf \Phi}$ the ($SO(2)$) SUSY invariant relations 
between $(\chi, B^i, \nu, F^i)$ and $\psi^i$ 
are constructed in all orders as follows: 
\ba
\A \A 
\chi = \xi^i \left[ \psi^i \vert w \vert 
+ {i \over 2} \kappa^2 \partial_a 
\{ \gamma^a \psi^i \bar\psi^j \psi^j 
(1 - i \kappa^2 \bar\psi^k \!\!\not\!\partial \psi^k) \} \right], 
\nonu
\A \A 
B^i = - \kappa \left( {1 \over 2} \xi^i \bar\psi^j \psi^j 
- \xi^j \bar\psi^i \psi^j \right) \vert w \vert, 
\nonu
\A \A 
\nu = \xi^i \epsilon^{ij} \left[ \psi^j \vert w \vert 
+ {i \over 2} \kappa^2 \partial_a 
\{ \gamma^a \psi^j \bar\psi^k \psi^k 
(1 - i \kappa^2 \bar\psi^l \!\!\not\!\partial \psi^l) \} \right], 
\nonu
\A \A 
F^i = {1 \over \kappa} \xi^i \left\{ \vert w \vert 
+ {1 \over 8} \kappa^4 
\Box ( \bar\psi^j \psi^j \bar\psi^k \psi^k ) 
\right\} 
- i \kappa \xi^j \partial_a ( \bar\psi^i \gamma^a \psi^j \vert w \vert ). 
\label{SSUSYinv}
\ea

Substituting Eqs.(\ref{VSUSYinv}) and (\ref{SSUSYinv}) 
into the free actions (\ref{Vaction}) and (\ref{Saction}) 
gives the equivalence of $S_{V0}$ and $S_{\Phi 0}$ to the VA action $S_{\rm VA}$ 
of Eq.(\ref{VAaction}) up to surface terms, respectively, i.e. 
\be
S_{\rm VA} = S_{V0} + [ \ {\rm surface\ terms} \ ] 
\ee
in all orders of $\psi^i$ \cite{ST2} when $\xi^2 = 1$, while 
\be
S_{\rm VA} = S_{\Phi 0} + [ \ {\rm surface\ terms} \ ] 
\label{VA-S0}
\ee
at least up to ${\cal O}(\kappa^0)$ when $(\xi^i)^2 = 1$. 
(We can expect that from the experiences the relation (\ref{VA-S0}) 
in all orders of $\psi^i$.) 
Therefore, the VA action is also equivalent to the sum of those free actions, 
$S_{V0} + S_{\Phi 0}$, as 
\be
S_{\rm VA} = S_{V0} + S_{\Phi 0} + [ \ {\rm surface\ terms} \ ], 
\label{VA-V0S0}
\ee
provided that $\xi^2 + (\xi^i)^2 = 1$. 

Now let us construct a LSUSY and gauge invariant interacting action 
by introducing a gauge interaction term 
for the $N = 2$ vector supermultiplet ${\bf V}$ 
and the $N = 2$ scalar supermultiplet ${\bf \Phi}$. 
We take the following interacting action 
where $e$ is a gauge coupling constant whose dimension is $({\rm mass})^1$, 
\ba
S_e = \A \A \int d^2 x \left[ 
e \left\{ i v_a \bar\chi \gamma^a \nu 
- \epsilon^{ij} v^a B^i \partial_a B^j 
+ \bar\lambda^i \chi B^i 
+ \epsilon^{ij} \bar\lambda^i \nu B^j 
- {1 \over 2} D (B^i)^2 \right. \right. 
\nonu
\A \A 
\left. \left. 
+ {1 \over 2} (\bar\chi \chi + \bar\nu \nu) A 
- \bar\chi \gamma_5 \nu \phi \right\}
+ {1 \over 2} e^2 (v_a{}^2 - A^2 - \phi^2) (B^i)^2 \right], 
\label{gaction}
\ea
and the total action, 
\be
S = S_{V0} + S'_{\Phi 0} + S_e, 
\label{totaction}
\ee
where $S'_{\Phi 0}$ means the LSUSY action (\ref{Saction}) 
without the FI terms $-{1 \over \kappa} \xi^i F^i$ due to the gauge invariance. 
The action (\ref{totaction}) is invariant 
under the LSUSY transformations (\ref{VLSUSY}) for ${\bf V}$ 
and the following LSUSY (gauge) transformations for ${\bf \Phi}$ 
suggested from the superfield technology 
$\delta_\zeta {\bf \Phi} \sim e {\bf V} {\bf \Phi}$ for $N = 1$ \cite{WB}, 
\ba
\delta_\zeta \chi 
\A = \A (F^i - i \!\!\not\!\partial B^i) \zeta^i - e \epsilon^{ij} V^i B^j, 
\nonu
\delta_\zeta B^i 
\A = \A \bar\zeta^i \chi - \epsilon^{ij} \bar\zeta^j \nu, 
\nonu
\delta_\zeta \nu 
\A = \A \epsilon^{ij} (F^i + i \!\!\not\!\partial B^i) \zeta^j + e V^i B^i, 
\nonu
\delta_\zeta F^i 
\A = \A - i \bar\zeta^i \!\!\not\!\partial \chi 
- i \epsilon^{ij} \bar\zeta^j \!\!\not\!\partial \nu 
\nonu
\A \A 
- e \{ \epsilon^{ij} \bar V^j \chi - \bar V^i \nu 
+ (\bar\zeta^i \lambda^j + \bar\zeta^j \lambda^i) B^j 
- \bar\zeta^j \lambda^j B^i \} 
\label{SLSUSYg}
\ea
with $V^i = i v_a \gamma^a \zeta^i - \epsilon^{ij} A \zeta^j - \phi \gamma_5 \zeta^i$. 
Eqs.(\ref{SLSUSYg}) satisfy the off-shell commutator algebra 
depending on $e$, 
\ba
\A \A 
[ \delta_{\zeta_1}, \delta_{\zeta_2} ] \chi = \Xi^a \partial_a \chi - e \theta \nu, 
\nonu
\A \A 
[ \delta_{\zeta_1}, \delta_{\zeta_2} ] B^i = \Xi^a \partial_a B^i - e \epsilon^{ij} \theta B^j, 
\nonu
\A \A 
[ \delta_{\zeta_1}, \delta_{\zeta_2} ] \nu = \Xi^a \partial_a \nu + e \theta \chi, 
\nonu
\A \A 
[ \delta_{\zeta_1}, \delta_{\zeta_2} ] F^i = \Xi^a \partial_a F^i + e \epsilon^{ij} \theta F^j, 
\label{N2D2commDI}
\ea
where $\theta$ is the generator of the $U(1)$ gauge transformation 
in Eq.(\ref{N2D2commg}). 
The closure of the commutator algebra (\ref{N2D2commDI}) 
as gauge covariant transformations of the fields can be seen 
in a manifest gauge invariant formulation of the LSUSY theory 
as explains below. 

Here let us show explicitly the gauge invariance of the action (\ref{totaction}). 
We define a complex (Dirac) spinor field $\chi_D$ 
and complex scalar fields $(B, F)$ in ${\bf \Phi}$ by 
\be
\chi_D = {1 \over \sqrt{2}} (\chi + i \nu), 
\ \ \ B = {1 \over \sqrt{2}} (B^1 + i B^2), 
\ \ \ F = {1 \over \sqrt{2}} (F^1 - i F^2), 
\label{cfields}
\ee
and express the $S'_{\Phi 0} + S_e$ in the action (\ref{totaction}) 
in terms of $(\chi_D, B, F)$ as 
\ba
S'_{\Phi 0} + S_e 
= \A \A \int d^2 x \{ i \bar\chi_D \!\not\!\!{\cal D} \chi_D 
+ \vert {\cal D}_a B \vert^2 + \vert F \vert^2 
\nonu
\A \A 
+ e (\bar\chi_D \lambda B + \bar\lambda \chi_D B^* - D \vert B \vert^2 
+ \bar\chi_D \chi_D A + i \bar\chi_D \gamma_5 \chi_D \phi) 
\nonu
\A \A 
- e^2 (A^2 + \phi^2) \vert B \vert^2 \} + [ \ {\rm surface\ term} \ ], 
\ea
where $\lambda = {1 \over \sqrt{2}} (\lambda^1 - i \lambda^2)$ 
and the covariant derivative ${\cal D}_a = \partial_a - i e v_a$. 
Then the invariance of the action (\ref{totaction}) 
under the ordinary local $U(1)$ gauge transformations, 
\ba
\A \A 
(\chi_D, B, F) \ \ \rightarrow \ \ (\chi'_D, B', F')(x) = e^{i \theta(x)} (\chi_D, B, F)(x), 
\nonu
\A \A 
v_a \ \ \rightarrow \ \ v'_a(x)  = v_a(x) + {1 \over e} \partial_a \theta(x), 
\ea
is manifest. 
The commutor algebra (\ref{N2D2commDI}) is also rewritten 
for the fields (\ref{cfields}) as 
\be
[ \delta_{Q1}, \delta_{Q2} ] = \delta_g({\cal D}), 
\label{N2D2commDII}
\ee
where $\delta_g({\cal D})$ means a gauge covariant transformation 
according to ${\cal D} = \Xi^a \partial_a + i e \theta$. 

Now we discuss the gauge interaction terms (\ref{gaction}) in NLSUSY representation 
based on the SUSY invariant relations (\ref{VSUSYinv}) and (\ref{SSUSYinv}) . 
Substituting Eqs.(\ref{VSUSYinv}) and (\ref{SSUSYinv}) into Eq.(\ref{gaction}) 
gives the following nonvanishing four fermion self-interaction terms, 
\be
S_e(\psi) \equiv \int d^2 x \left\{ 
{1 \over 4} e \kappa \xi (\xi^i)^2 \bar\psi^j \psi^j\bar\psi^k \psi^k 
\right\}. 
\label{4fermion}
\ee
(The terms proportional to $e^2$ vanish due to $(\psi^i)^5 \equiv 0$.) 
In order to reduce the action (\ref{totaction}) to the VA action (\ref{VAaction}) 
from the NLSUSYGR (SGM \cite{KS}) scenario for everything, 
the terms (\ref{4fermion}) corresponding to the action (\ref{gaction}) 
should be cancelled out by other four fermion self-interaction terms, 
e.g. those which appear from other parts of Eq.(\ref{totaction}) 
in NLSUSY representation. 
For this purpose we redefine the SUSY invariant relations (\ref{SSUSYinv}) 
for the scalar supermultiplet ${\bf \Phi}$, 
in particular, we add four fermion self-interaction terms 
depending on the gauge coupling constant $e$ 
to the original $F^i = F^i(\psi)$ as follows: 
\ba
\tilde F^i = \A \A {1 \over \kappa} \xi^i \left\{ \vert w \vert 
+ {1 \over 8} \kappa^3 
\Box ( \bar\psi^j \psi^j \bar\psi^k \psi^k ) 
\right\} 
- i \kappa \xi^j \partial_a ( \bar\psi^i \gamma^a \psi^j \vert w \vert ) 
\nonu
\A \A 
- {1 \over 4} e \kappa^2 \xi \xi^i \bar\psi^j \psi^j \bar\psi^k \psi^k. 
\label{newSUSYinv}
\ea
The new set of SUSY invariant relations is denoted 
by ${\bf \tilde \Phi} = {\bf \tilde \Phi}(\psi)$, 
which corresponds to $(\chi(\psi), B^i(\psi), \nu(\psi), \tilde F^i(\psi))$ 
in Eqs.(\ref{SSUSYinv}) and (\ref{newSUSYinv}), 
while the original set of the SUSY invariant relations (\ref{VSUSYinv}) 
for the vector supermultiplet are maintained and denoted as ${\bf V} = {\bf V}(\psi)$ below. 
By substituting the ${\bf \tilde \Phi}(\psi)$ and ${\bf V}(\psi)$ into Eq.(\ref{totaction}), 
the $S(\psi)$ reduces to the $N = 2$ VA action (\ref{VAaction}) 
by means of the cancellations among the four fermion self-interaction terms as follows: 
\ba
S(\psi) \A = \A (S_{V0} + S'_{\tilde \Phi 0} + S_e)(\psi) 
\nonu
\A \equiv \A S_{\rm VA} + \int d^2 x \left\{ 
- {1 \over 4} e \kappa \xi (\xi^i)^2 \bar\psi^k \psi^k \bar\psi^l \psi^l 
+ {1 \over 4} e \kappa \xi (\xi^i)^2 \bar\psi^k \psi^k \bar\psi^l \psi^l 
\right\} 
\nonu
\A \A 
+ [{\rm suface\ terms}] 
\nonu
\A = \A 
S_{\rm VA} + [{\rm surface\ terms}], 
\label{VA-gauge}
\ea
where the terms 
$- {1 \over 4} e \kappa \xi (\xi^i)^2 \bar\psi^k \psi^k \bar\psi^l \psi^l$ 
in the second line appear from ${1 \over 2} (\tilde F^i)^2$ in $S'_{\tilde \Phi 0}$. 
Now $S_{V0} + S'_{\tilde \Phi 0} + S_e$ is equivalent to $N = 2$ VA action 
provided $\xi^2 - (\xi^i)^2 = 1$ for the real parameters $\xi$ and $\xi^i$ 
in Eq.(\ref{VA-gauge}) due to the absence of the FI terms 
$-{1 \over \kappa} \xi^i \tilde F^i$ from $S'_{\tilde \Phi 0}$. 

In the LSUSY theory (\ref{VA-gauge}) equivalent to the VA model, 
the component fields $(\chi, B^i, \nu, \tilde F^i)$ in ${\bf \tilde \Phi}$ 
constitute the LSUSY scalar multiplet, 
where the LSUSY transformations are realized as Eqs.(\ref{SLSUSYg}) 
and their commutator algebra is closed as Eq.(\ref{N2D2commDI}) 
(or Eq.(\ref{N2D2commDII})). 
On the other hand, the variations of ${\bf \tilde \Phi} = {\bf \tilde \Phi}(\psi)$ 
by means of the NLSUSY transformations (\ref{NLSUSY}) 
can be computed straightforwardly, 
and by comparing these variations with the LSUSY transformations (\ref{SLSUSYg}) 
for the component fields in ${\bf \tilde \Phi}$, 
they (except $\delta_\zeta B^i(\psi)$) are recasted as follows: 
\ba
\delta_\zeta \chi(\psi) 
\A = \A ({\rm LSUSY}(\ref{SLSUSYg}))(\psi) + X(\zeta; \psi), 
\nonu
\delta_\zeta \nu(\psi) 
\A = \A ({\rm LSUSY}(\ref{SLSUSYg}))(\psi) + Y(\zeta; \psi), 
\nonu
\delta_\zeta \tilde F^i(\psi) 
\A = \A ({\rm LSUSY}(\ref{SLSUSYg}))(\psi) + Z^i(\zeta; \psi), 
\label{SLSUSYgpII}
\ea
where (LSUSY(\ref{SLSUSYg}))($\psi$) means Eqs.(\ref{SLSUSYg}) 
in terms of ${\bf \tilde \Phi} = {\bf \tilde \Phi}(\psi)$ 
and ${\bf V} = {\bf V}(\psi)$. 
$(X, Y, Z^i)(\zeta; \psi)$ are defined as 
\ba
X(\zeta; \psi) 
\A = \A {1 \over 4} e \kappa^2 \xi \xi^i \bar\psi^j \psi^j \bar\psi^k \psi^k \zeta^i, 
\nonu
Y(\zeta; \psi) 
\A = \A {1 \over 4} e \kappa^2 \xi \xi^i \epsilon^{ij} \bar\psi^k \psi^k \bar\psi^l \psi^l \zeta^j, 
\nonu
Z^i(\zeta; \psi) \A = \A e \kappa \xi \{ 
\xi^j (\bar\zeta^i \psi^j - \bar\zeta^j \psi^i) \bar\psi^k \psi^k 
(1 - 2i \kappa^2 \bar\psi^l \!\!\not\!\partial \psi^l) 
\nonu
\A \A
- i \kappa^2 \xi^i \bar\zeta^j \psi^j \bar\psi^k \psi^k 
\bar\psi^l \!\!\not\!\partial \psi^l \}. 
\label{XYZ}
\ea
(Note that $- e \epsilon^{ij} V^i(\psi) B^j(\psi)$ and $e V^i(\psi) B^i(\psi)$ 
of $\delta_\zeta \chi(\psi)$ and $\delta_\zeta \nu(\psi)$ 
in Eq.(\ref{SLSUSYgpII}) vanish respectively.) 
Equation (\ref{SLSUSYgpII}) shows that the variations of 
${\bf \tilde \Phi} = {\bf \tilde \Phi}(\psi)$ 
by means of the NLSUSY transformations (\ref{NLSUSY}) 
become the sum of the LSUSY transformations (\ref{SLSUSYg}) 
for ${\bf \tilde \Phi}$ 
and $(X, Y, Z^i)(\zeta; \psi)$ defined as Eqs.(\ref{XYZ}). 

In the above arguments, 
we find that ${\bf \tilde \Phi} = {\bf \tilde \Phi}(\psi)$ 
is the set of {\it SUSY invariant} relations 
in a sense that the commutator algebra under the NLSUSY transformations (\ref{NLSUSY}) 
is closed as $[ \delta_{Q1}, \delta_{Q2} ] = \delta_P(\Xi^a)$ of Eq.(\ref{N2D2comm}). 
In particular, this is the case for the new $\tilde F^i = \tilde F^i(\psi)$ 
of Eqs.(\ref{newSUSYinv}) due to 
\be
[ \delta_{\zeta_1}, \delta_{\zeta_2} ] 
\left( - {1 \over 4} e \kappa^2 \xi \xi^i \bar\psi^j \psi^j \bar\psi^k \psi^k \right) 
= \Xi^a \partial_a 
\left( - {1 \over 4} e \kappa^2 \xi \xi^i \bar\psi^j \psi^j \bar\psi^k \psi^k \right). 
\label{comm4f}
\ee
Namely, in Eq.(\ref{VA-gauge}) 
the original definition of the SUSY invariant relations (\ref{SSUSYinv}) 
is relaxed to the ${\bf \tilde \Phi} = {\bf \tilde \Phi}(\psi)$ 
with respect to their variations under the NLSUSY transformations (\ref{NLSUSY}) 
as in Eqs.(\ref{SLSUSYgpII}) with the $(X, Y, Z^i)(\zeta; \psi)$. 
In appearance, these situations are similar to Eqs.(\ref{NLSUSYv}), (\ref{W}) 
and (\ref{N2D2comm}) for $\delta_\zeta v^a(\psi)$. 
Therefore, we conclude that the $N = 2$ VA action (\ref{VAaction}) 
is equivalent to the LSUSY and local U(1) gauge invariant action (\ref{totaction}) 
with the FI term, 
i.e. 
\be
S_{\rm VA} = S_{V0} + S'_{\tilde \Phi 0} + S_e + [{\rm surface\ terms}]. 
\label{VA-gauge:}
\ee
under the new definition of the SUSY invariant relations 
${\bf \tilde \Phi} = {\bf \tilde \Phi}(\psi)$ 
in addition to ${\bf V} = {\bf V}(\psi)$. 

The results obtained in this paper are summarized as follows. 
Firstly, we have linearized $N = 2$ NLSUSY VA model in $d = 2$ 
and shown by using the free SUSY invariant relations (\ref{VSUSYinv}) and (\ref{SSUSYinv}) 
it is equivalent to the LSUSY (free) theory which is described 
by $N = 2$ ($SO(2)$) vector and scalar supermultiplets as depicted in Eq.(\ref{VA-V0S0}). 
Next, we have constructed the (covariant) gauge interaction terms (\ref{gaction}) 
for these LSUSY multiplets and studied the relation 
between the VA action (\ref{VAaction}) 
and the LSUSY local $U(1)$ gauge invariant action (\ref{totaction}). 
Contrary to the vanishments of the NLSUSY representation 
of the Yukawa and the mass terms \cite{ST2} 
the LSUSY gauge interaction terms produce 
the nonvanishing four fermion self-interaction terms (\ref{4fermion}) 
when the free SUSY invariant relations (\ref{SSUSYinv}) and (\ref{VSUSYinv}) are adopted. 
We have shown that such a nontrivial (and harmful from the NLSUSYGR scenario) 
terms (\ref{4fermion}) are cancelled out in the action (\ref{totaction}) 
in the NLSUSY representation by means of the new SUSY invariant relations (\ref{SSUSYinv}), 
in particular, by means of the redefinition of the new $\tilde F^i(\psi)$ 
of Eq.(\ref{newSUSYinv}) where the four fermion self-interaction terms 
depending on the gauge coupling constant $e$ are added to the free case 
SUSY invariant relations $F^i(\psi)$. 
Then the equivalence of the action (\ref{totaction}) to the $N = 2$ VA action 
(\ref{VAaction}) has been shown as in Eq.(\ref{VA-gauge}), 
where the component fields in ${\bf \tilde \Phi}$ constitute the LSUSY scalar multiplet. 
For the proof of this equivalence we have changed (relaxed) the initial definition 
of the SUSY invariant relations (\ref{SSUSYinv}) valid for the free case 
into the new interacting case ${\bf \tilde \Phi} = {\bf \tilde \Phi}(\psi)$ 
with the consequent NLSUSY transformations (\ref{SLSUSYgpII}) 
containing the $(X, Y, Z^i)(\zeta; \psi)$. 
In spite of the relaxation, the closure of the commutator algebra 
for the ${\bf \tilde \Phi} = {\bf \tilde \Phi}(\psi)$ 
under the NLSUSY transformations (\ref{NLSUSY}) has been proved as Eq.(\ref{N2D2comm}), 
in particular, for the new $\tilde F^i(\psi)$ due to Eq.(\ref{comm4f}). 
Therefore, we have concluded in Eq.(\ref{VA-gauge:}) that the $N = 2$ VA model 
is equivalent to the spontaneously broken $N = 2$ LSUSY gauge theory 
with the action (\ref{totaction}) 
in a sense that the new definition of the SUSY invariant relations 
${\bf \tilde \Phi} = {\bf \tilde \Phi}(\psi)$ including Eqs.(\ref{newSUSYinv}) holds. 

Since the (free) SUSY invariant relations (\ref{VSUSYinv}) 
for the vector supermultiplet ${\bf V}$ are maintained in the linearization, 
the LSUSY Yukawa interaction terms 
$S_{Vf} = \int d^2 x 
\{ f ( A \bar\lambda^i \lambda^i + \epsilon^{ij} \phi \bar\lambda^i \gamma_5 \lambda^j 
+ A^2 D - \phi^2 D - \epsilon^{ab} A \phi F_{ab} ) \}$ 
and the mass terms 
$S_{Vm} = \int d^2 x 
\left\{ - {1 \over 2} m \ 
( \bar\lambda^i \lambda^i - 2 AD + \epsilon^{ab} \phi F_{ab} ) \right\}$ 
in ${\bf V}$ \cite{ST2} 
can be added to the equivalent LSUSY gauge theory without violating the equivalence 
(\ref{VA-gauge:}). 
Namely,
\be
S_{\rm VA} = S_{V0} + S'_{\tilde \Phi 0} + S_e + S_{Vf} + S_{Vm} 
+ [{\rm suface\ terms}] 
\ee
holds for ${\bf V} = {\bf V}(\psi)$ 
and ${\bf \tilde \Phi} = {\bf \tilde \Phi}(\psi)$ 
as one possible configulation of the basic component fields, 
which is equivalent to the $N = 2$ VA model. 
Once the gauge and Yukawa interaction terms, $S_e$ and $S_{Vf}$, 
are added to the free action $S'_{\tilde \Phi 0}$ and 
$S_{V0}$ containing the FI term, they become nontrivial in the action, 
break SUSY spontaneously 
and produce mass for the composite fields automatically 
as demonstrated in Ref.\cite{ST4}. 
These results are very favourable for the SGM \cite{KS} scenario 
where all particles except graviton are the composite 
of NG-fermion (called superons) of the NLSUSY VA model. 
It is interesting that the four-fermion self-interaction term 
(i.e. the condensation of $\psi^i$) which contributes to the auxiliary field 
is the origin of the familiar local $U(1)$ gauge symmetry of LSUSY theory. 
As for the SUSY QED equivalence of $N = 2$ VA model in $d = 4$, 
the similar results are anticipated but further investigations are needed.

%
%

\newpage

%
\newcommand{\NP}[1]{{\it Nucl.\ Phys.\ }{\bf #1}}
\newcommand{\PL}[1]{{\it Phys.\ Lett.\ }{\bf #1}}
\newcommand{\CMP}[1]{{\it Commun.\ Math.\ Phys.\ }{\bf #1}}
\newcommand{\MPL}[1]{{\it Mod.\ Phys.\ Lett.\ }{\bf #1}}
\newcommand{\IJMP}[1]{{\it Int.\ J. Mod.\ Phys.\ }{\bf #1}}
\newcommand{\PR}[1]{{\it Phys.\ Rev.\ }{\bf #1}}
\newcommand{\PRL}[1]{{\it Phys.\ Rev.\ Lett.\ }{\bf #1}}
\newcommand{\PTP}[1]{{\it Prog.\ Theor.\ Phys.\ }{\bf #1}}
\newcommand{\PTPS}[1]{{\it Prog.\ Theor.\ Phys.\ Suppl.\ }{\bf #1}}
\newcommand{\AP}[1]{{\it Ann.\ Phys.\ }{\bf #1}}

\end{document}